\documentclass[pdftex,twocolumn,epjc3]{svjour3}      
\RequirePackage[T1]{fontenc}

\smartqed  

\RequirePackage{graphicx}
\RequirePackage{mathptmx}      
\RequirePackage{flushend}
\RequirePackage[switch]{lineno}
\RequirePackage[numbers,sort&compress]{natbib}
\RequirePackage[colorlinks,citecolor=blue,urlcolor=blue,linkcolor=blue]{hyperref}

\journalname{Eur. Phys. J. C}
\RequirePackage{orcidlink}
\RequirePackage{multirow}
\RequirePackage{array}
\RequirePackage{amsmath}
\begin{document}

\title{Pulse shape discrimination for $\alpha$ event rejection in BEGe-type high-purity germanium detectors}

\author{A. Biondi \thanksref{1,2,e1} \orcidlink{0009-0009-9199-7814}, K. Szczepaniec \thanksref{2,e2} \orcidlink{0000-0003-3229-777X}
        \and
        T. Mróz \thanksref{2,e3} \orcidlink{0000-0002-5304-5531}
        \and
        M. Misiaszek \thanksref{2} \orcidlink{0000-0001-5726-9666}
        \and
        G. Zuzel \thanksref{2} \orcidlink{0000-0001-5898-2658}
}

\thankstext{e1}{Corresponding Author: \href{mailto:alex.biondi@doctoral.uj.edu.pl}{alex.biondi@doctoral.uj.edu.pl}}
\thankstext{e2}{Presently at INFN Laboratori Nazionali del Gran Sasso, Assergi (AQ), 67100, Italy}
\thankstext{e3}{Presently at the Henryk Niewodniczański
Institute of Nuclear Physics, Polish Academy of Sciences, ul. Radzikowskiego 152, 31-342 Krakow, Poland}

\institute{Jagiellonian University, Doctoral School of Exact and Natural Sciences, ul. Łojasiewicza 11, 30–348 Kraków, Poland\label{1} 
\and M. Smoluchowski Institute of Physics, Jagiellonian University, ul. Łojasiewicza 11, 30–348 Kraków, Poland \label{2} 
} 

\date{Received: date / Accepted: date}

\maketitle

\begin{abstract} 
High-purity germanium detectors are widely used in rare-event searches due to their excellent energy resolution and extremely high intrinsic (radio)purity. In experiments searching for neutrinoless double beta decay in $^{76}$Ge such as \textsc{Legend}, pulse shape discrimination is required to suppress multi-site $\gamma$ events. In this work, we investigate whether pulse shape discrimination classifiers trained exclusively on $\gamma$ ray data can be used to identify and reject $\alpha$ events, without the need for dedicated $\alpha$ training. In detectors such as \textsc{Legend}, the total number of registered $\alpha$ events over the experiment lifetime is expected to be insufficient to train dedicated classifiers, while still contributing to the background. Two approaches based on machine learning are studied: a multilayer perceptron and a projective likelihood classifier. The p+ surface of a point-contact semi-planar germanium detector was exposed to $^{209}$Po and $^{210}$Po sources deposited on a thin gold foil. Two measurement campaigns were performed, yielding $1.36\times10^{5}$ and $1.87\times10^{6}$ $\alpha$ events, respectively.
Both classification methods achieve efficient separation of single-site and multi-site $\gamma$ events while strongly reducing the $\alpha$ component. The multilayer perceptron provides the best overall performance, with a signal-like event survival greater than 80\%, a background-like event survival below 20\%, and an $\alpha$-rejection factor exceeding $2.71\times10^{4}$. These results demonstrate that robust pulse shape discrimination for high-purity germanium detectors can be achieved using training information derived solely from $\gamma$ events, providing a promising strategy for next-generation neutrinoless double beta decay searches.
\end{abstract}

\section{Introduction}
\label{sec:intro}

High-purity germanium (HPGe) detectors have been successfully applied in $\gamma$-spectrometry due to their excellent energy resolution and high detection efficiency. In addition, they are also deployed in searches for rare events, such as dark matter particles interactions or neutrinoless double beta ($0\nu\beta\beta$) decay. The latter can theoretically occur in several isotopes, leading to multiple experiments investigating this hypothetical transition in different candidate elements, such as $^{82}\mathrm{Se}$ \cite{cupid_se}, $^{100}\mathrm{Mo}$ \cite{nemo_mo,amore_mo,cupid_mo}, $^{130}\mathrm{Te}$ \cite{cuore_te} or $^{136}\mathrm{Xe}$ \cite{kamland_xe,exo_xe}, among others.
The $0\nu\beta\beta$ decay is also expected in $^{76}\mathrm{Ge}$, which motivates the use of HPGe detectors enriched in this isotope. Their excellent energy resolution, low intrinsic background and the fact that they can serve as both the source and the detector at the same time (maximizing the detection efficiency \cite{GERDA_FINAL}) make them an interesting option for the target material. These features reduce the overall complexity of the experimental setup and help to minimize the background, enabling a quasi background-free operation. Previous experiments, such as Heidelberg–Moscow \cite{HdM_FINAL}, \textsc{Igex} \cite{IGEX_FINAL}, \textsc{Majorana Demonstrator} \cite{MAJORANA_FINAL} and \textsc{Gerda} \cite{GERDA_FINAL}, have systematically pushed the limit of the achievable background index (BI) and have improved the measured lower limit on the $0\nu\beta\beta$ half-life. \textsc{Legend}~\cite{legend2025first} has taken over the effort, building on the success of \textsc{Gerda} and \textsc{Majorana Demonstrator}.

The \textsc{Gerda} collaboration has reported a BI of $5.2^{+1.6}_{-1.3} \times 10^{-4}\ \mathrm{counts}/(\mathrm{keV}\cdot\mathrm{kg}\cdot\mathrm{yr})$ \cite{GERDA_FINAL}, which is one of the lowest among the previous generation of $0\nu\beta\beta$ experiments. Performing a frequentist analysis, the lower limit on the $0\nu\beta\beta$ half-life in $^{76}\mathrm{Ge}$ was estimated to be $T_{1/2} > 1.8 \times 10^{26}$ years at a $90\%$ confidence level \cite{GERDA_FINAL}. \textsc{Legend} has recently presented the results of its first year of data taking, showing a background level slightly lower than the one achieved by \textsc{Gerda}. From the combined analysis of data from \textsc{Gerda}, \textsc{Majorana Demonstrator} and \textsc{Legend-200}, it has set the most stringent limit on the $0\nu\beta\beta$ half-life in $^{76}\mathrm{Ge}$ of $T_{1/2} > 1.9 \times 10^{26}$ years at a $90\%$ confidence level \cite{legend2025first}.

\begin{figure}
\centering
\includegraphics[width=0.50\textwidth]{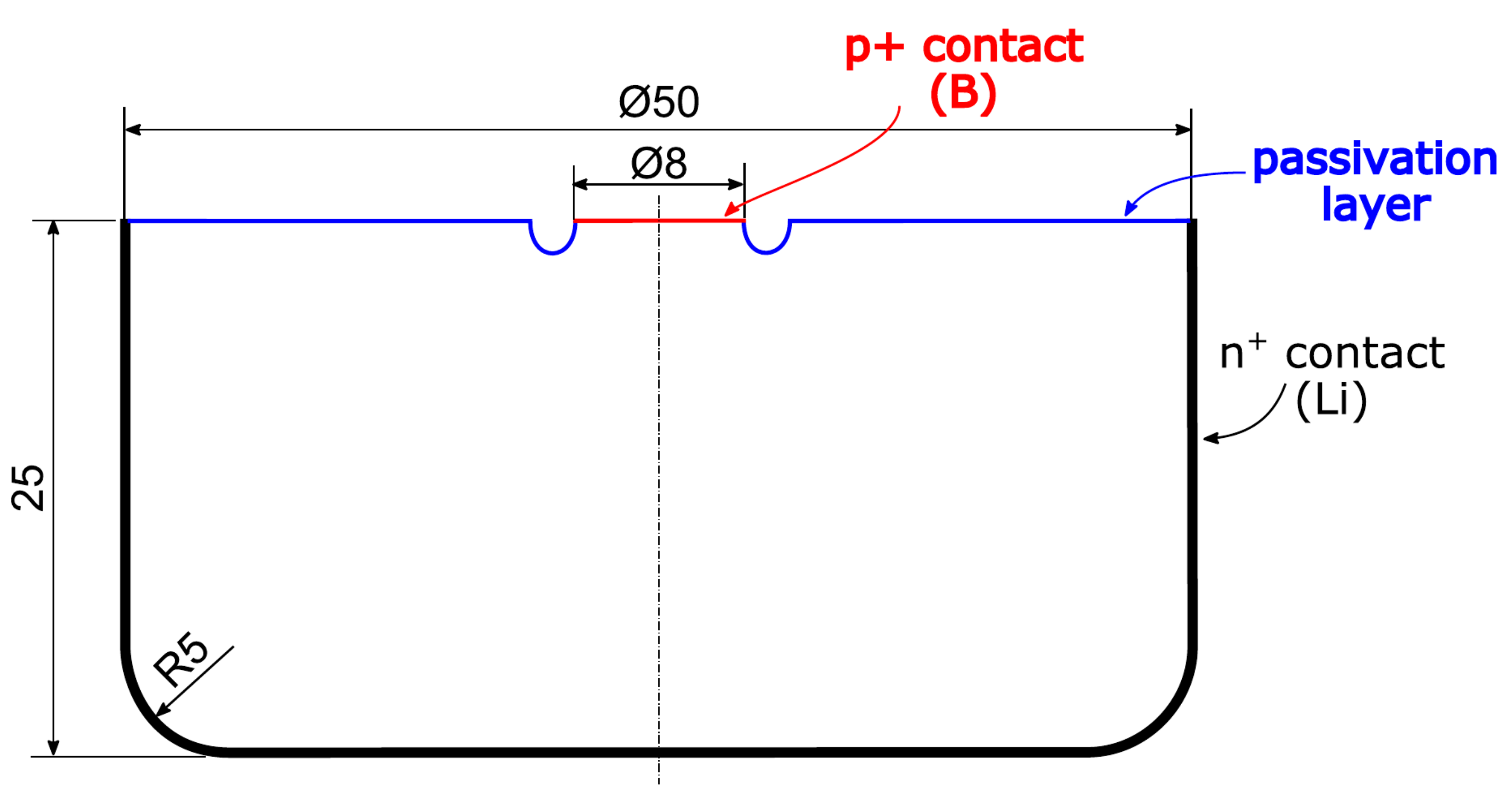}
\caption{Schematic of the HPGe detector used in this study. The thick black line represents the Li-diffused $n+$ contact. The red and blue lines represent the B-implanted $p+$ electrode and the passivated surface, respectively. Dimensions are given in mm. Figure taken from \cite{Jany2021}.}
\label{fig:bege}
\end{figure}

In experiments searching for rare events, a significant source of background may originate from the long-lived daughters of $^{222}\mathrm{Rn}$, namely $^{210}\mathrm{Po}$ and $^{210}\mathrm{Pb}$. Although the latter does not contribute directly, it decays with a half-life of 22.3 yr and ultimately produces $^{210}\mathrm{Po}$ through the short-lived intermediate $^{210}\mathrm{Bi}$, thereby generating a persistent background contribution.
 Contamination with other alpha emitters such as $^{226}\mathrm{Ra}$, which is present in dust, may also be relevant. If present on detector surfaces, the $\alpha$ particles (emitted by $^{210}\mathrm{Po}$) can penetrate the active region of the detectors and deposit either almost their full energy or only a fraction of it (after losing energy in the inactive layers), creating a detectable signal. A full energy deposition is usually not problematic in $0\nu\beta\beta$ decay experiments as it is relatively far from their region of interest (ROI). Contamination of the $n+$ contact, which is formed by thermally diffused Li (see Fig.~\ref{fig:bege}) and has a thickness of $\sim$ 0.7 mm, can be neglected. The $\alpha$ particles are typically stopped by the thick dead layer before reaching the active volume, as the range of 5.3 MeV $\alpha$ in Ge is $\sim 20$ $\mu$m. However, if the $\alpha$ emitter is sitting directly on the thin $p+$ contact (formed by implanted B, $O(0.1)$ $\mu$m thick), a significant portion of the deposited energy can be recorded. Therefore, such $\alpha$ particles may contribute to the events registered in the ROI of the experiment. The $0\nu\beta\beta$ signal in $^{76}\mathrm{Ge}$ is expected at $Q_{\beta\beta} = 2039.06 \pm 0.01$ keV \cite{GERDA_FINAL}. The most prominent $\alpha$ emission of $^{210}\mathrm{Po}$ has an energy of $E = 5304$ keV, and under suitable conditions (i.e., emission at a nonzero angle relative to the detector surface), the registered energy of such an $\alpha$ particle can be close to $Q_{\beta\beta}$.

\begin{figure}
\centering
\includegraphics[width=0.45\textwidth]{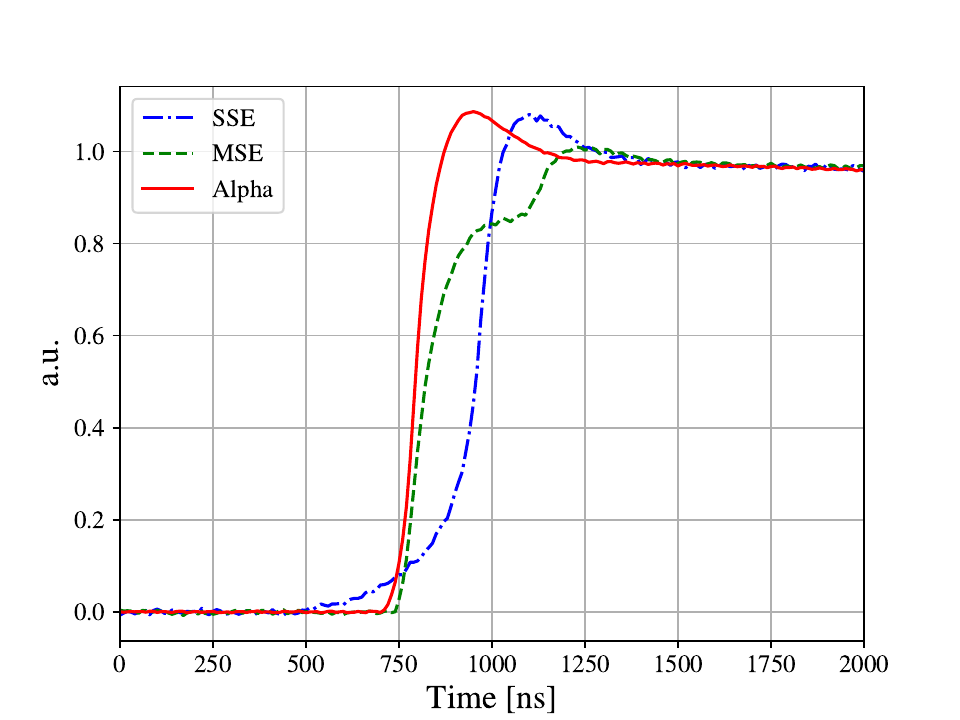}
    \caption{\label{fig:sampleevent}
    Example of real single-site, multi-site and alpha waveforms as recorded by the HPGe detector used in this study using an FADC card. Pulses were sampled with $100$ MHz frequency.}
\end{figure}

Pulse shape discrimination (PSD) methods can be applied at the offline analysis stage to eliminate such unwanted events \cite{agostini2013psd,agostini2022psd,alvis2019psd}. The basic concept of PSD is to differentiate between signals originating from the detector on a waveform basis and to reject undesirable events. In HPGe detectors, two primary classes of events can be identified: single-site events (SSEs) and multi-site events (MSEs). This distinction naturally arises from the multiplicity of energy depositions within the detector bulk.    
A $\gamma$ ray can deposit energy at a single location within the crystal, for instance through a single Compton scattering, producing an SSE. However, if it undergoes multiple Compton scatterings at separated locations within the crystal, it produces a MSE. These two classes can be distinguished by the shape of the corresponding waveforms, which reflect how the charge clouds are collected from the detector bulk. Example waveforms for both classes are shown in Fig.~\ref{fig:sampleevent}, together with a typical alpha event. For the $0\nu\beta\beta$ decay in $^{76}\mathrm{Ge}$, the energy deposition is expected to occur within a volume of about $1\,\mathrm{mm}^3$, producing an SSE. This property makes all MSE events an unwanted background by definition. Various PSD methods have already been demonstrated to be effective against $\gamma$ ray events. The \textsc{Gerda} and \textsc{Legend} collaborations have utilized the so-called $A/E$ classifier \cite{agostini2022psd}. Artificial neural networks have also been shown to effectively reject $\gamma$ ray multi-site events while preserving a large fraction of single-site events \cite{Misiaszek2018,Jany2021}. An $\alpha$ ray event typically deposits all its energy in less than 50 $\mu$m of germanium layer, therefore it is also SSE-like. However, the fact that the event takes place close to the p+ contact leads to a very fast charge collection, which may, in principle, induce differences in the pulse shapes compared to SSE.

However, there is no comprehensive study in the literature on the rejection of $\alpha$ ray events using PSD due to the practical difficulty of obtaining a clean sample. Furthermore, the number of such $\alpha$ events is very small in low-background experiments, making them insufficient for the development of effective cuts on their own. In contrast, regular calibration runs with well-known $\gamma$ emitters, for example $^{228}\mathrm{Th}$ as used in the \textsc{Legend} experiment, provide a high-statistics sample of clean $\gamma$ ray events suitable for PSD training.
The aim of this work is to test whether a PSD classifier trained and calibrated solely on $\gamma$ ray events can also effectively eliminate $\alpha$ ray events without additional adaptation. A satisfactory performance for a given PSD method would include:  
1) a high survival probability for single-site $\gamma$ ray events together with a high rejection probability for multi-site events, comparable with those obtained by \textsc{Gerda} \cite{GERDA_FINAL} and \textsc{Legend} \cite{legend2025first}, and  
2) the ability to achieve the highest possible suppression of $\alpha$ ray events.
We tested two PSD methods based on machine learning: the projective likelihood estimator (PLK) and the multilayer perceptron (MLP) artificial neural network (ANN). For comparison, we also considered the $A/E$ classifier.

\section{\label{sec:experimental_setup}Experimental setup}

In this study, we used a point-contact germanium detector, commonly referred to as a BEGe-type, shown schematically in Fig.~\ref{fig:bege}. The manufacturing process and detector characterization are described in detail in \cite{Jany2021}. The detector was installed in a conventional vacuum cryostat operated at surface level and shielded by 10 cm of lead. An additional plastic scintillator was placed above the shielding to veto cosmic muons. Detector pulses were recorded directly from the preamplifier using a 100 MHz, 14-bit FADC card. For energy calibration, a $^{228}\mathrm{Th}$ source was placed inside the lead shield, directly on top of the detector end-cap. The recorded $\gamma$ events were also used as input for the investigated PSD methods.

To obtain a sufficient number of $\alpha$ ray events, a dedicated $\alpha$ source was prepared in the form of a gold foil with polonium deposited on its surface. The foil was then placed directly on top of the $p+$ contact of the detector. The diameter of the gold foil was $d = 4$~mm, smaller than the physical diameter of the $p+$ contact ($D = 8$~mm). This ensured that almost all the $\alpha$ rays emitted toward the detector were injected into the $p+$ contact, allowing for a controlled study of the PSD response to $\alpha$ events occurring specifically on the p+ contact, while minimizing the contribution of slower signals from the groove or passivated surfaces. In our first attempt, $^{209}\mathrm{Po}$ was deposited on the gold foil, while in the second attempt we used $^{210}\mathrm{Po}$. Details of the polonium deposition procedure on gold are given in ~\ref{sec:source_po209}. The JFET transistor was replaced following a breakdown between the first and second campaign. As the electronics affects the PSD response of the detector, the PSD performance before and after the replacement of the JFET may not be comparable directly.

\section{Experimental Method}
\label{sec:exp_method}

\begin{figure}
\centering
\includegraphics[width=0.45\textwidth]{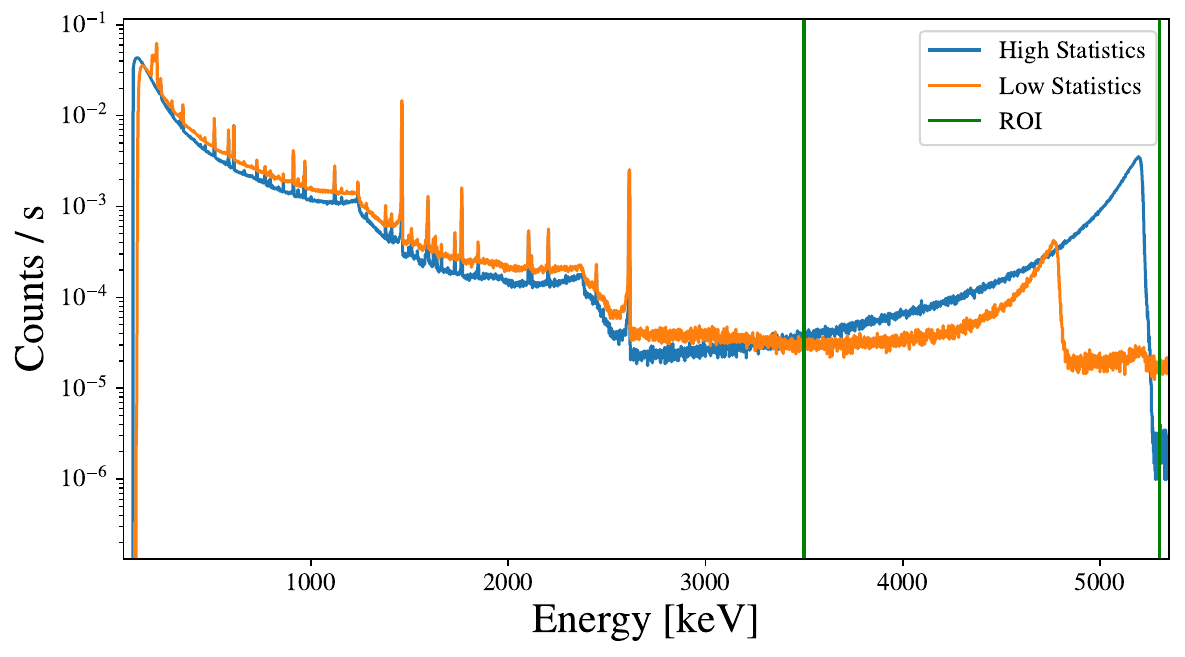}
\caption{Spectra of the high statistics (blue) and low statistics (orange) physics runs. The ROI is defined as the region 3500 keV < E < 5300 keV, and is indicated by the green vertical lines.}
\label{fig:spectra_bef} 
\end{figure}

Two measurement campaigns were carried out on the surface: the first using $^{209}\mathrm{Po}$ and the second using $^{210}\mathrm{Po}$, with source activities of $\sim$ 0.1 $\mathrm{Bq}$ and $\sim$ 1 $\mathrm{Bq}$, respectively. $^{209}\mathrm{Po}$ decays predominantly via the emission of a $4.88$ MeV $\alpha$ particle to $^{205}\mathrm{Pb}$, with a branching ratio of approximately $99.5\%$, while the remaining $0.5\%$ undergoes electron capture to $^{209}\mathrm{Bi}$; $^{210}\mathrm{Po}$ decays to $^{206}\mathrm{Pb}$ via the emission of a $5.30$ MeV $\alpha$ particle. For simplicity, the two measurement campaigns are referred to as the low statistics (LS) and high statistics (HS) campaigns, respectively. For each campaign, three datasets were collected: a calibration run with $^{228}\mathrm{Th}$, a physics run with the $\alpha$ source only, and a background run without any source. In this study, muons from cosmic rays are the dominant background contribution above 3.5 MeV. The calibration runs contain about 25 million waveforms (LS) and 90 million waveforms (HS). For the LS campaign, the calibration data used for PSD training were collected with the $\alpha$ source installed in the detector, whereas for the HS run they were collected after its removal. The livetimes of the physics runs are 826~h and 1122~h, while those of the background runs are 446~h and 2007~h for the LS and HS campaigns, respectively.
Events identified as unsuitable (noise, early or late triggers, pile-ups, negative leading edges, etc.) were removed through quality cuts. The muon veto and the germanium detector are connected to two channels in the same acquisition card, with the trigger set on the germanium detector. If any signal is registered by the scintillator within the DAQ acquisition window, the germanium event is rejected.
Baseline subtraction was applied to the pulses, and a trapezoidal filter was then used to extract their energy in ADC units. The energy scale was fixed using the $^{228}\mathrm{Th}$ data.
The ROI for this study was defined as the energy interval [3500, 5300]~keV to avoid contributions from $\gamma$ peaks and to include the $\alpha$ peaks and tails from both $^{209}\mathrm{Po}$ and $^{210}\mathrm{Po}$ isotopes. The upper bound of the ROI is sufficient to fully contain the observed $^{210}\mathrm{Po}$ peak within the detector energy resolution, since the $\alpha$ particles do not deposit their full nominal energy in the active volume. Even for perpendicular emission, a small fraction of the $\alpha$-particle energy is lost in the p+ contact before reaching the active region, shifting the reconstructed $^{210}\mathrm{Po}$ peak below its nominal energy. Within the ROI, $2.18 \times 10^{5}$ and $1.89 \times 10^{6}$ events remained after all quality and veto cuts, for the LS and HS runs, respectively. Spectra of the HS and LS physics runs with the ROI highlighted are shown in Fig.~\ref{fig:spectra_bef}. After subtracting the background counts from the physics data in the ROI, we estimate that $1.36 \times 10^{5}$ and $1.87 \times 10^{6}$ events, respectively, originate from $\alpha$ particles.

The analysis is performed using the raw waveforms as input.
For the machine learning (ML) algorithms, we select 64 points belonging to the rising edge of the waveform: 32 samples before, 31 after, and one at the maximum of the current signal. This dimensionality reduction preserves the information-rich portion of the waveform while significantly reducing the number of input features \cite{Misiaszek2018}. A standard scaler, fitted on the training dataset, is then applied to the selected features. Two ML models were tested: the PLK classifier and the MLP ANN.
The projective likelihood classifier belongs to the family of Naive Bayesian classifiers and is therefore based on the product of the likelihoods of the individual input features. General information on Naive Bayesian classifiers can be found in \cite{zhang2004bayesian,Bishop2006bayesian}. For each input variable and for each class, a one-dimensional probability density is constructed from the training data using binned distributions smoothed by cubic-spline interpolation and normalized to unit area. Assuming statistical independence among the input variables, the class likelihood is obtained by combining the corresponding one-dimensional densities. Posterior class probabilities are then derived by including class priors and performing the computation in logarithmic form to ensure numerical stability. Although in our case the input features are not statistically independent, the projective likelihood approach can still provide a useful and stable classifier. This is a common property of Naive Bayesian methods, which often perform well in classification tasks despite violations of the independence assumption.
MLPs are widely used in physics for non-linear classification tasks \cite{Misiaszek2018,Jany2021}. They consist of multiple layers of interconnected artificial neurons whose connection strengths, or weights, are determined during training. Non-linear activation functions provide the flexibility required to model complex decision boundaries, while the feed-forward structure ensures efficient evaluation once the network is trained \cite{prince2023deeplearning}. 
In this work, the architecture we used is composed of six dense layers of artificial neurons with dimensions [64, 32, 16, 8, 4, 1]. The first five layers are followed by a ReLU activation function, while a sigmoid activation function was used in the output layer. L1 regularization was applied to the weights of the first layer. The architecture was chosen empirically to balance model complexity with the available training data size, while hyperparameters (such as learning rate and batch size) were optimized with a grid search. The Adam optimizer was used for training. No additional dropout layers were included.

\begin{figure}
\centering
\includegraphics[width=0.50\textwidth]{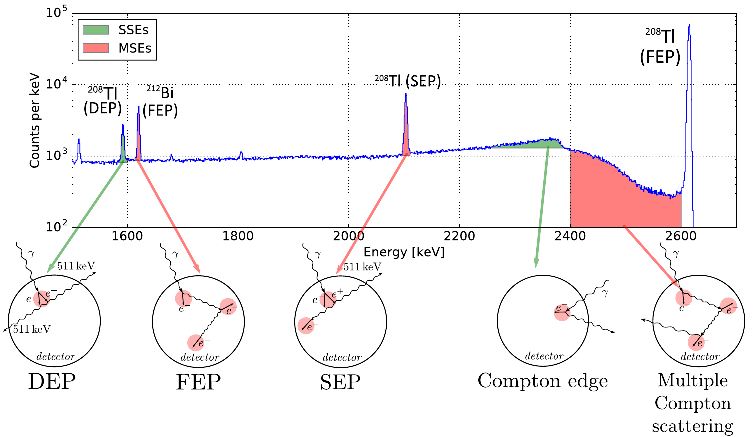}
\caption{$^{228}$\textrm{Th} spectrum. The regions dominated by SSEs are highlighted in green, while those dominated by MSEs are highlighted in pink. Figure taken from \cite{Misiaszek2018}.}
\label{fig:thorium_psd} 
\end{figure}

\begin{figure}
\centering
\includegraphics[width=0.45\textwidth]{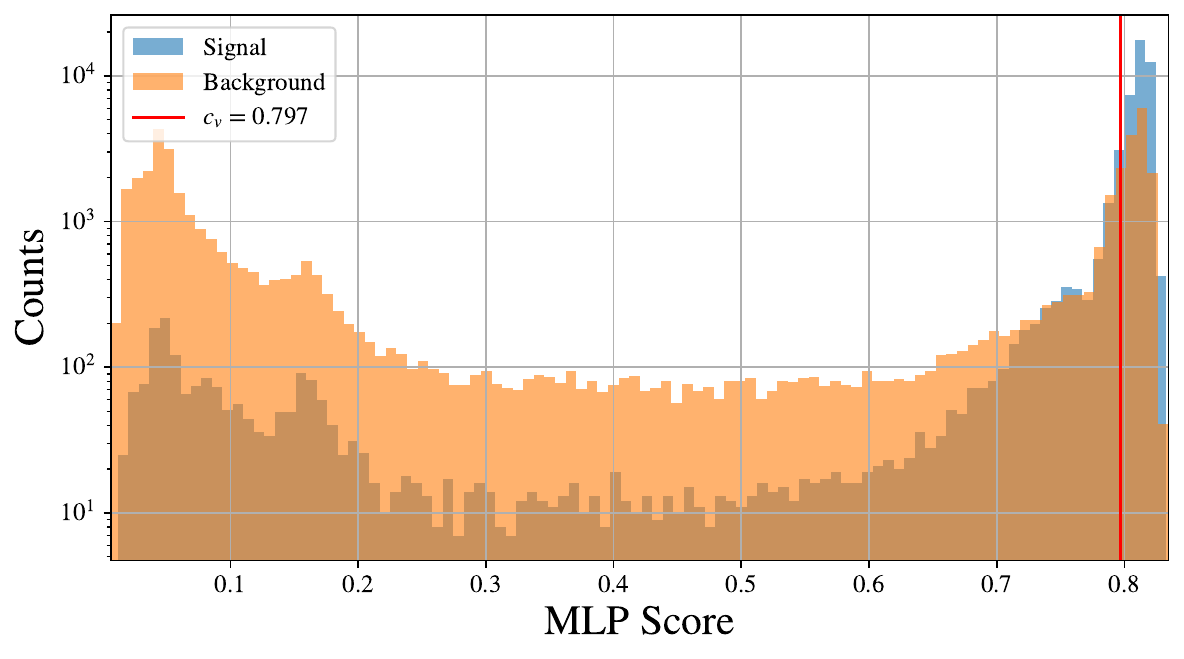}
\caption{Example of the distribution of the MLP predictions for the HS training dataset with the TC2 configuration. The signal (SSE) distribution is shown in blue, while the background (MSE) distribution is shown in orange. The red line represents the selected cut. Plots for other configurations listed in Tab. \ref{tab:training_config} are analogous.}
\label{fig:cut_selection} 
\end{figure}

The training process is based on a calibration run with $^{228}\mathrm{Th}$, whose spectrum contains several regions that provide relatively clean samples of SSE and MSE events. Specifically, the double escape peak (DEP) and Compton edge (CE) are characteristic of SSEs, while the single escape peaks (SEPs), full energy peaks (FEPs), and the multi-Compton scattering (MCS) region are representative of MSEs. The regions of the $^{228}\mathrm{Th}$ spectrum used for PSD training are highlighted in Fig. \ref{fig:thorium_psd} \cite{Misiaszek2018}. We selected six configurations to train our models, listed in Tab. \ref{tab:training_config}.
To ensure the purity of the training sample, only the central part of each peak is selected, corresponding to approximately $E_0 \pm0.5$ FWHM for $^{208}\mathrm{Tl}$ SEP and $^{212}\mathrm{Bi}$ FEP, and $E_0 \pm0.2$ FWHM for $^{208}\mathrm{Tl}$ DEP. For the CE, the energy region [2370,2380] keV was chosen for the training, while for the MCS the energy region [2450,2550] keV was used. The training dataset consists of approximately 20 000 and 100 000 events per configuration for the LS and HS runs, respectively. To avoid biases, the dataset was balanced so that it contained the same number of SSE and MSE events.
An example of the distribution of the MLP model predictions for the training dataset with the TC2 configuration is shown in Fig. \ref{fig:cut_selection}; the distributions for the remaining configurations listed in Tab. \ref{tab:training_config} are analogous.

In addition, $A/E$ classifier is applied for comparison. The $A/E$ parameter is defined as the ratio between the maximum of the current signal and the energy of the event in ADC counts. The $E$ parameter is the energy extracted by the trapezoidal filter. The waveform is differentiated to obtain the current pulse, and a one-dimensional Gaussian filter is used to filter out the noise. The maximum of the filtered current pulse is the $A$ parameter. The $A/E$ distribution is analyzed using the routines provided in the \texttt{pygama} \cite{agostini_pygama} Python package developed by the \textsc{Legend} Collaboration. After correcting the $A/E$ distribution for its energy dependence and normalizing it to the mean value of the DEP of $^{208}\mathrm{Tl}$, the $A/E$ classifier is calculated \cite{agostini2022psd}.


\begin{table}

\centering
\caption{Configurations used to train the PLK and MLP classifiers}

\begin{tabular}{lll}

\hline\noalign{\smallskip}

\textbf{Configuration} & \textbf{SSE} & \textbf{MSE}  \\
\noalign{\smallskip}\hline\noalign{\smallskip}
TC1 & $^{208}$\textrm{Tl} DEP & $^{208}$\textrm{Tl} SEP \\
TC2 & $^{208}$\textrm{Tl} DEP & $^{212}$\textrm{Bi} FEP \\
TC3 & $^{208}$\textrm{Tl} DEP & $^{208}$\textrm{Tl} MCS \\
TC4 & $^{208}$\textrm{Tl} CE & $^{208}$\textrm{Tl} SEP \\
TC5 & $^{208}$\textrm{Tl} CE & $^{212}$\textrm{Bi} FEP \\
TC6 & $^{208}$\textrm{Tl} CE & $^{208}$\textrm{Tl} MCS \\
\noalign{\smallskip}\hline
\end{tabular}
\label{tab:training_config}
\end{table}

To select the cut value $c_v$, we consider as a figure of merit (FoM) the significance of a counting experiment \cite{Cowan2015fom,Vischia2019FoM}, defined as the ratio between the signal and the square root of the sum of signal and background:

\begin{equation}
    \mathrm{FoM}'(x) = \frac{S(x)}{\sqrt{S(x) + B(x)}}
\end{equation}

where $x$ represents a possible value of the cut, and $S$ and $B$ denote the signal and background counts. In our case, the signal is chosen to be the $^{208}\mathrm{Tl}$ DEP, which serves as a proxy for the $0\nu\beta\beta$ decay, while the background is represented by the MSE peaks of the $^{208}\mathrm{Tl}$ SEP and FEP. Defining $\varepsilon$ as the efficiency (or survival fraction) of the PSD cut, we can write:
\begin{gather*}
    S(x) = \varepsilon_{DEP}(x)\,S_{0\,DEP} \\ 
    B(x) = \varepsilon_{SEP}(x)\,B_{0\,SEP} + \varepsilon_{FEP}(x)\,B_{0\,FEP}
\end{gather*}

where $S_0$ and $B_0$ indicate the number of counts in each peak before applying the PSD cut. Substituting these expressions and dividing by $\sqrt{S_0}$, which is an overall constant factor, the FoM can be rewritten as:

\begin{equation}
        \mathrm{FoM}(x) = \frac{\varepsilon_{DEP}(x)}{\sqrt{\varepsilon_{DEP}(x) + \frac{B_{0\,SEP}}{S_{0\,DEP}}\,\varepsilon_{SEP}(x) + \frac{B_{0\,FEP}}{S_{0\,DEP}}\,\varepsilon_{FEP}(x) }}
\end{equation}

\begin{figure}
\centering
\includegraphics[width=0.45\textwidth]{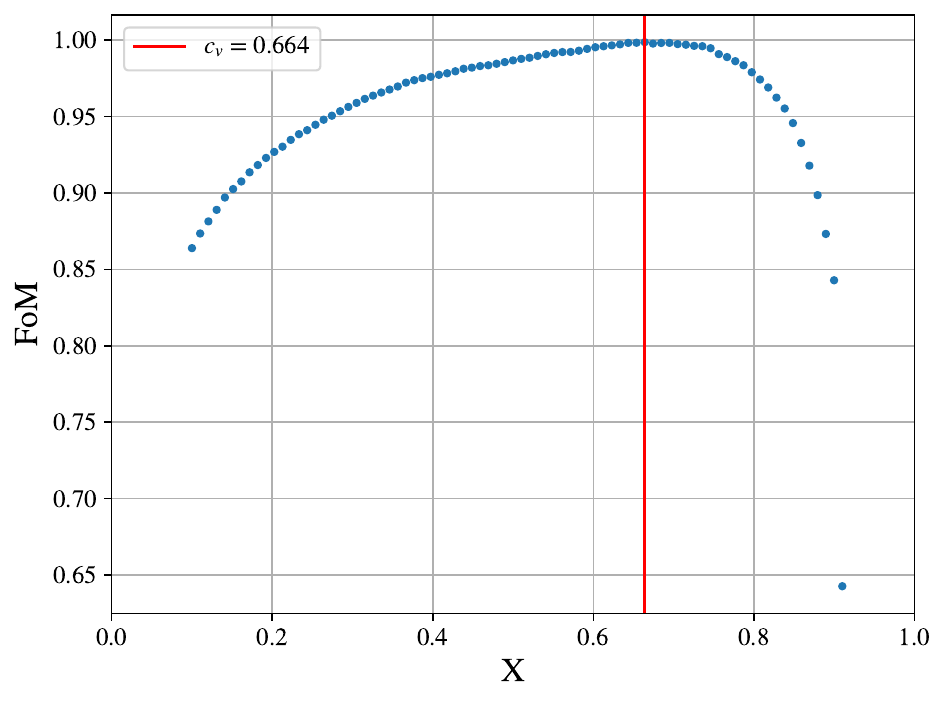}
\includegraphics[width=0.45\textwidth]{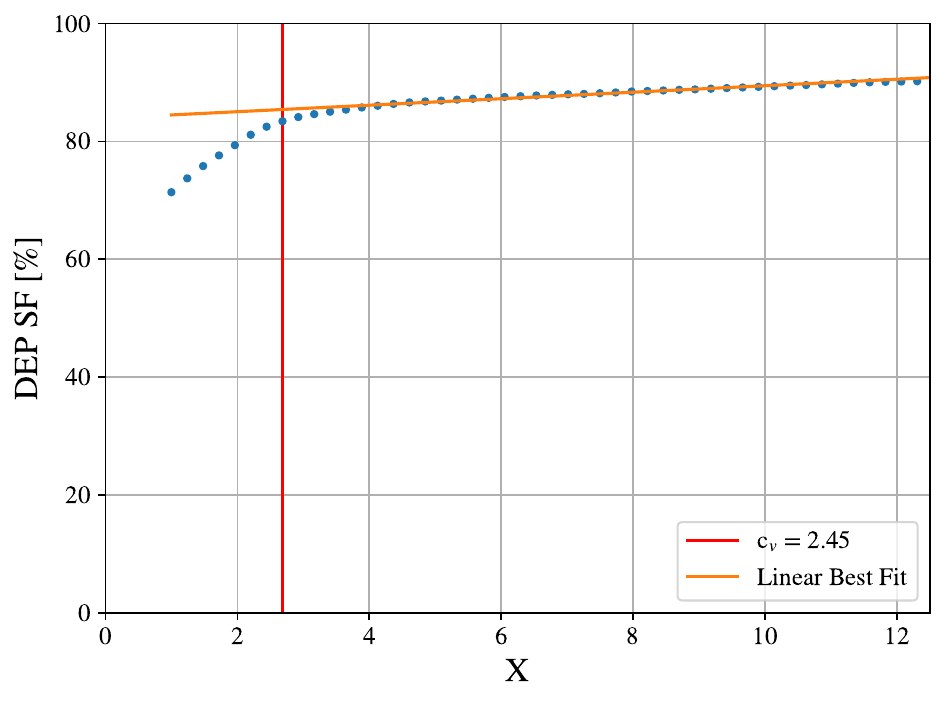}

\caption{Example of the distribution of the FoM for the MLP in the LS run (top). Example of selection the $A/E$ high cut for the LS run (bottom). The red lines correspond to the selected cut values.}
\label{fig:fom} 
\end{figure}

The efficiencies are obtained by performing two Gaussian fits to each peak, one before and one after applying the PSD cut. This approach accounts for the possible reduction of the Compton continuum. In fact, the peak area is defined as:
\begin{equation*}
    A = \sqrt{2\pi}\cdot N \sigma
\end{equation*}

Where $N$ is the amplitude of the Gaussian and $\sigma$ its standard deviation. The uncertainties on the survival fractions are propagated from the statistical uncertainties of the fits.
For the ML algorithms, the PSD cut is chosen as the value that maximizes the FoM. An example of the FoM distribution and the corresponding cut selection is shown in Fig. \ref{fig:fom} (top). For the $A/E$ classifier, the lower cut is chosen to retain 90\% of DEP events, while the upper cut is selected at the point where the DEP survival fraction begins to decrease rapidly with increasing cut value, as shown in Fig. \ref{fig:fom} (bottom).

\begin{figure}
\centering
\includegraphics[width=0.45\textwidth]{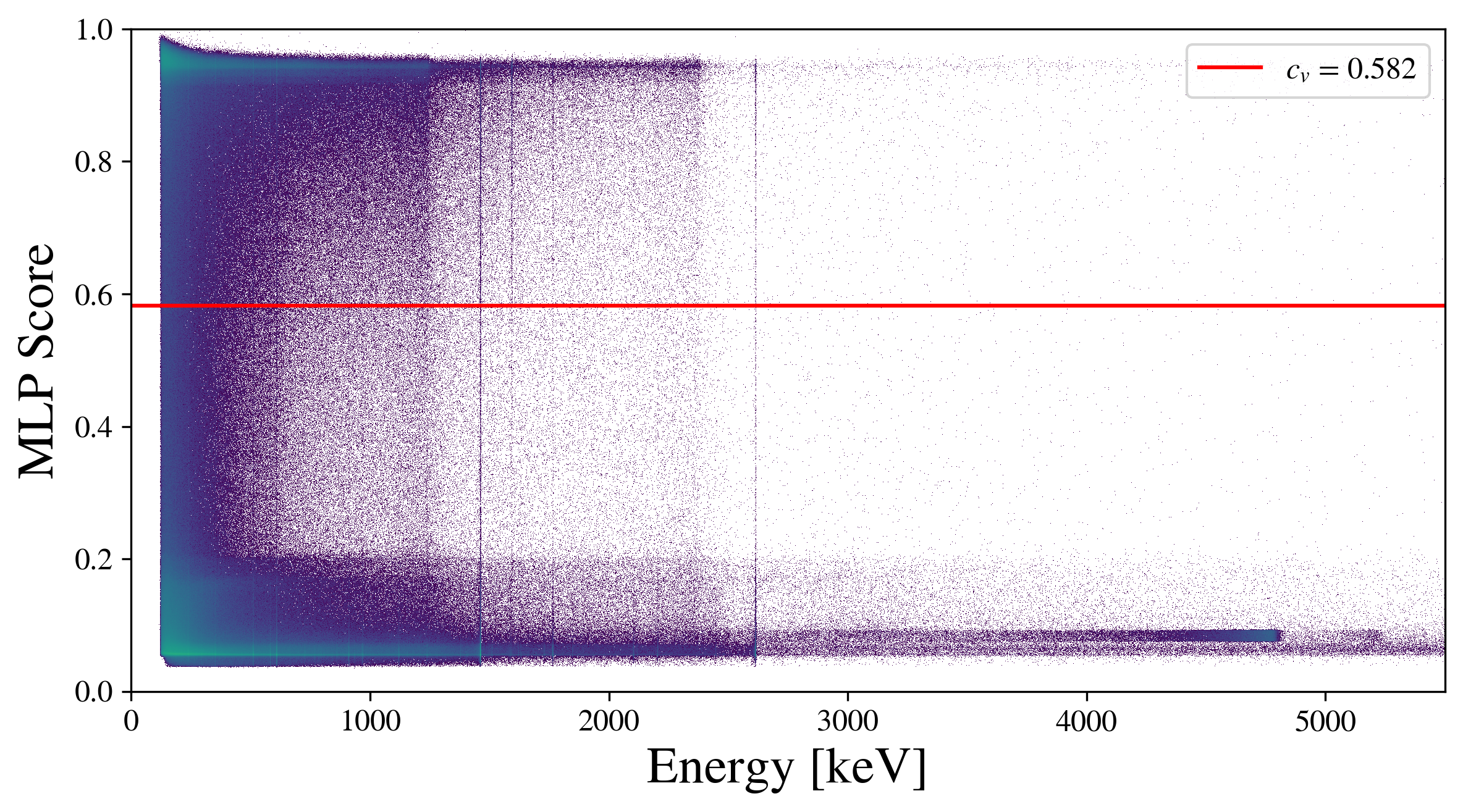}
\includegraphics[width=0.45\textwidth]{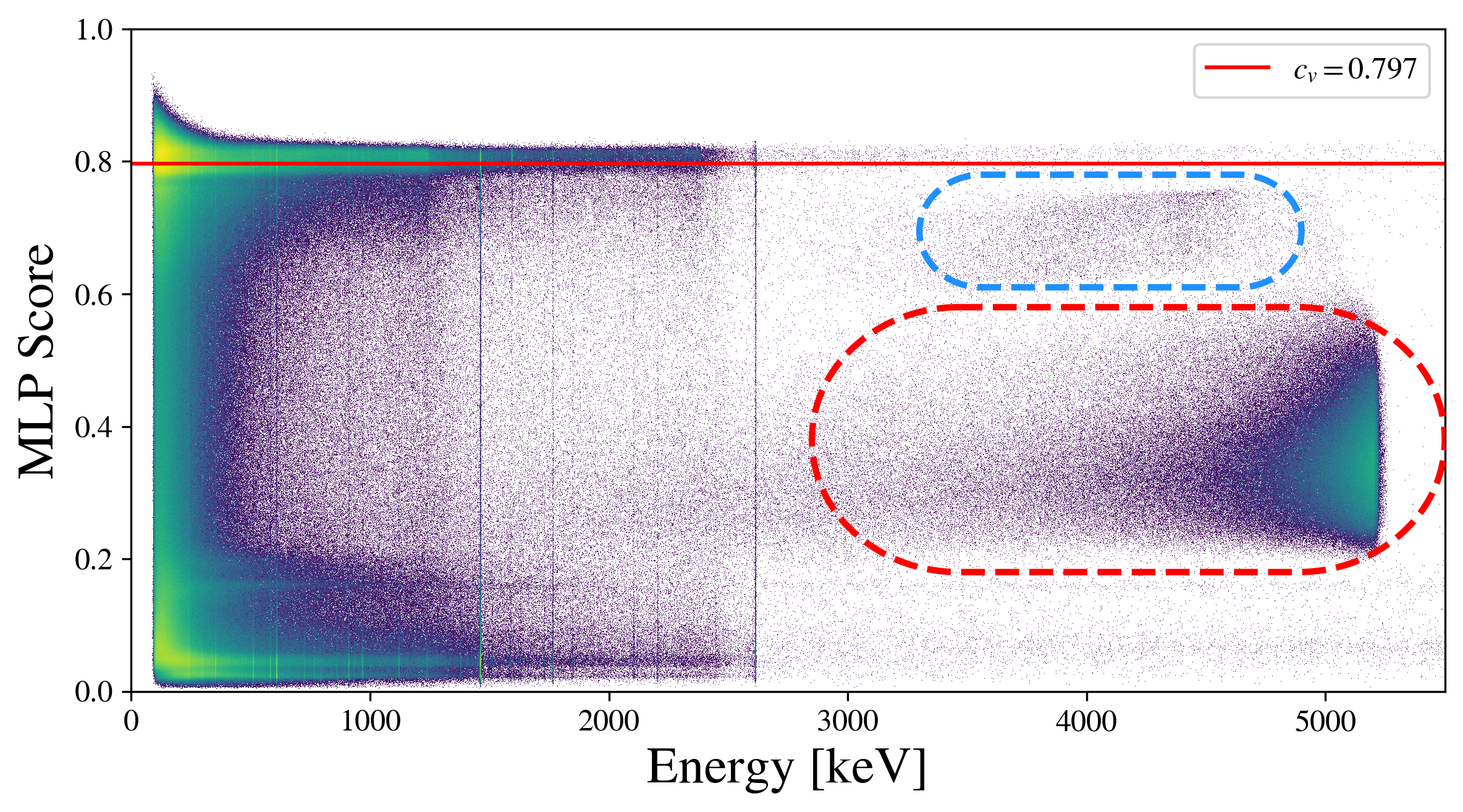}
\caption{MLP score vs energy scatter plots for the physics runs (LS top, HS bottom). The color intensity is proportional to the counts in each bin. The red horizontal line represents the selected cut. The two dashed shapes highlight the two alpha clusters present in the HS run.}
\label{fig:scatter_example} 
\end{figure}

\begin{figure}
\centering
\includegraphics[width=0.45\textwidth]{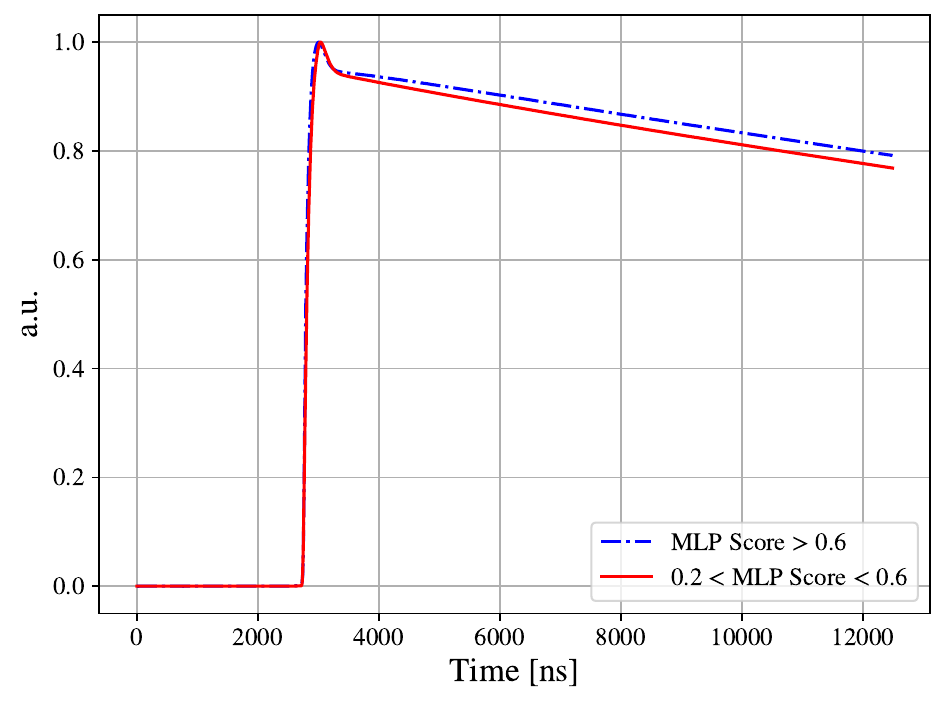}
    \caption{\label{fig:alphas_wf}
    Average pulses from the two alpha clusters present in the HS run.}
\end{figure}

The distribution of the model predictions as a function of energy is then plotted for all runs. An example of the scatter plots for the physics runs is presented in Fig.~\ref{fig:scatter_example}.
For the LS run, the $\alpha$ particles are confined in a region close to an MLP score of 0.2, which lies significantly below the cut. The rest of the high-energy events in the spectrum are due to cosmic rays.
For the HS run, most of the $\alpha$ events form a large cluster between MLP scores of 0.2 and 0.6. However, a second cluster, with energies between 3500 keV and 4800 keV, lies considerably closer to the signal band, even though it is still below the cut value that maximizes the FoM. The two clusters are highlighted with dashed shapes in Fig. \ref{fig:scatter_example} (bottom).
The average pulse shapes of the two clusters are compared in Fig. \ref{fig:alphas_wf}. The presence of the cluster with an MLP score above 0.6 might be related to the non-flatness of the source, which may generate some events near the edges of the contact.
Notably, this cluster is absent in the LS run, which may be explained by the fact that such events are rare and therefore negligible in the lower $\alpha$ statistics conditions expected for the $0\nu\beta\beta$ decay experiments. Indeed, in the HS run they represent only $\sim$ 1\% of the $\alpha$ events.

\begin{figure}
\centering
\includegraphics[width=0.50\textwidth]{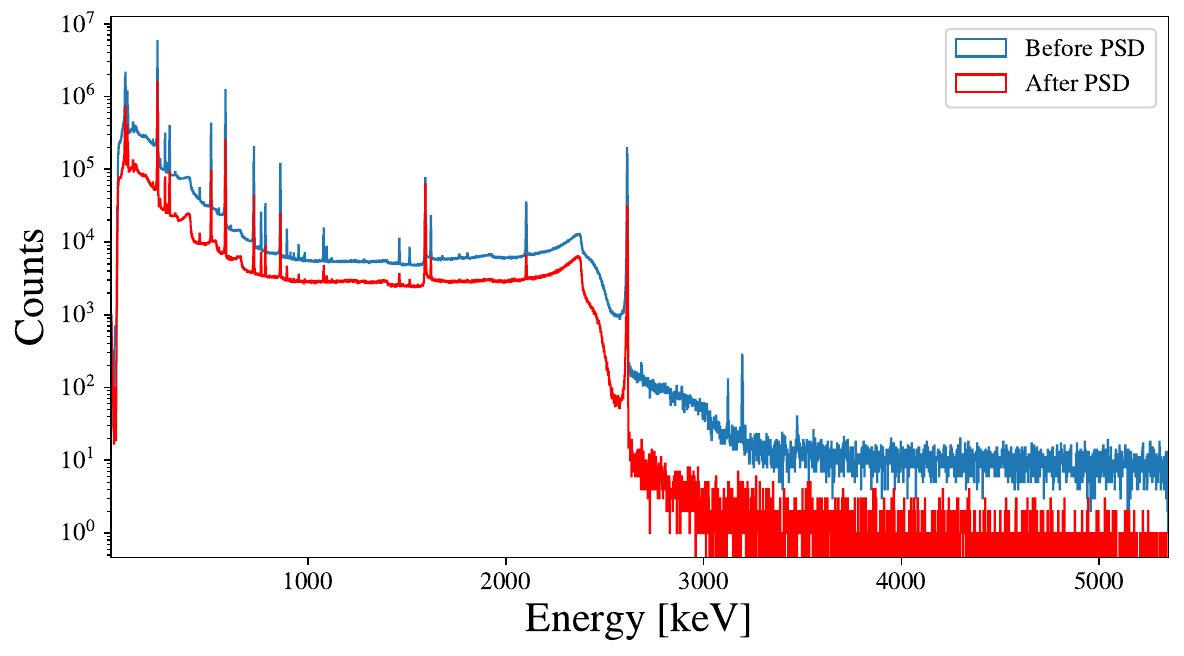}
\caption{Spectrum of the HS calibration run before and after the application of the PSD cut with MLP.}
\label{fig:spectrum_cal_aft}
\end{figure}

The chosen cut value $c_v$ is then applied to the calibration, physics, and background runs and the corresponding spectra are produced. An example of the spectrum of the HS calibration run before and after PSD is shown in Fig.~\ref{fig:spectrum_cal_aft}. To evaluate the performance of the models, we consider the survival fractions of the signal and background peaks across the $\gamma$ spectrum, as well as the rejection factor for the $\alpha$ events.
For the calculation of the $\alpha$-rejection factor, the number of counts in the ROI before applying the PSD cut is first determined for both the physics and background runs. Since the runs have different durations in both the LS and HS campaigns, the number of counts in the longer run is scaled to match the livetime of the shorter one. The two values are then statistically subtracted to obtain the number of $\alpha$ events before the PSD cut. The same procedure is applied to the counts in the ROI after the cut. If the resulting value is negative or compatible with zero within its uncertainty, a 90\% confidence level limit is assigned. The $\alpha$-rejection factor is then calculated as the ratio between the number of $\alpha$ events before and after the PSD cut. In cases where the background-subtracted counts after the cut represent only an upper limit, the corresponding rejection factor is quoted as a lower limit.
An example of this statistical procedure is reported in Tab.~\ref{tab:statistical_analysis}.

\begin{table}[htbp]
\caption{Example of the statistical analysis performed to calculate the alpha rejection factor for the HS run with the TC2 configuration. The numbers of counts are evaluated within the ROI.}
\label{tab:statistical_analysis}
\centering
\scriptsize
\begin{tabular}{lll}
\hline
 & \textbf{Physics Run} & \textbf{Background Run} \\
\hline
Time [s] & 4 042 561 & 7 225 813 \\
Counts before PSD & 1.89$\times$10$^6 \pm$ 1376 & 3.9$\times$10$^4 \pm$ 197 \\
Counts before PSD normalized & 1.89$\times$10$^6 \pm$ 1376 & 2.18$\times$10$^4 \pm$ 110 \\
Counts after PSD & 1846 $\pm$ 43 & 3370$\pm$ 58 \\
Counts after PSD normalized & 1846 $\pm$ 43 & 1885 $\pm$ 32 \\
\hline
\multicolumn{3}{l}{\textbf{Background-subtracted quantities:}} \\
Counts before PSD & \multicolumn{2}{c}{1.87$\times$10$^6 \pm$ 1381} \\
Counts after PSD & \multicolumn{2}{c}{$-39 \pm 54$} \\
Counts after PSD 90\% C.L. & \multicolumn{2}{c}{$< 69$} \\
\textbf{Alpha Rejection Factor (90\% C.L.)} & \multicolumn{2}{c}{\textbf{$> 2.71 \times 10^4$}} \\
\hline
\end{tabular}
\end{table}

\section{Results}
\label{sec:results}

\begin{table*}[htbp]
\centering
\caption{Survival fractions for the high statistics dataset (MLP configurations TC1--TC6 and A/E benchmark).}
\label{tab:results_HS_MLP}
\begin{tabular}{c c *{6}{c} c}
\hline
\multirow{2}{*}{Energy [keV]} & 
\multirow{2}{*}{Isotope} &
\multicolumn{7}{c}{Survival Fraction [\%]} \\
\cline{3-9}
 & & TC1 & TC2 & TC3 & TC4 & TC5 & TC6 & A/E \\
\hline
727.3  & $^{212}$Bi       & 19.9 $\pm$ 0.2 & 20.0 $\pm$ 0.2 & 21.3 $\pm$ 0.2 & 40.0 $\pm$ 0.5 & 30.2 $\pm$ 0.3 & 46.3 $\pm$ 0.5 & 40.5 $\pm$ 0.4 \\
763.0  & $^{208}$Tl       & 20.2 $\pm$ 0.7 & 20.4 $\pm$ 0.7 & 21.5 $\pm$ 0.7 & 37.1 $\pm$ 0.8 & 28.8 $\pm$ 0.7 & 48.3 $\pm$ 0.9 & 40.2 $\pm$ 0.8 \\
785.0  & $^{212}$Bi       & 19.5 $\pm$ 0.4 & 19.8 $\pm$ 0.4 & 21.1 $\pm$ 0.4 & 37.2 $\pm$ 0.5 & 28.4 $\pm$ 0.4 & 47.1 $\pm$ 0.6 & 38.8 $\pm$ 0.5 \\
860.6  & $^{208}$Tl       & 18.2 $\pm$ 0.2 & 19.1 $\pm$ 0.2 & 19.9 $\pm$ 0.2 & 36.9 $\pm$ 0.3 & 31.0 $\pm$ 0.3 & 46.0 $\pm$ 0.4 & 37.1 $\pm$ 0.3 \\
1078   & $^{212}$Bi       & 18 $\pm$ 1     & 18 $\pm$ 1     & 18 $\pm$ 1     & 34 $\pm$ 1     & 31 $\pm$ 1     & 44 $\pm$ 1     & 34 $\pm$ 1 \\
1093   & $^{208}$Tl       & 14 $\pm$ 3     & 15 $\pm$ 3     & 15 $\pm$ 3     & 39 $\pm$ 4     & 26 $\pm$ 3     & 39 $\pm$ 4     & 27 $\pm$ 3 \\
1512   & $^{212}$Bi       & 17 $\pm$ 2     & 16 $\pm$ 2     & 18 $\pm$ 2     & 29 $\pm$ 3     & 30 $\pm$ 2     & 41 $\pm$ 3     & 32 $\pm$ 2 \\
\textbf{1592} & \textbf{$^{208}$Tl (DEP)} 
        & $\mathbf{83.3 \pm 0.8}$ & $\mathbf{84.0 \pm 0.8}$ & $\mathbf{79.8 \pm 0.8}$ & $\mathbf{84.0 \pm 0.8}$ & $\mathbf{84.1 \pm 0.8}$ & $\mathbf{87 \pm 1}$ & $\mathbf{73.6 \pm 0.7}$ \\
1620.5 & $^{212}$Bi       & 17.3 $\pm$ 0.8 & 17.2 $\pm$ 0.6 & 18.5 $\pm$ 0.6 & 28.1 $\pm$ 0.9 & 30.6 $\pm$ 0.9 & 45 $\pm$ 1     & 33.0 $\pm$ 0.8 \\
2103   & $^{208}$Tl (SEP) & 10.1 $\pm$ 0.3 & 11.8 $\pm$ 0.3 & 13.2 $\pm$ 0.3 & 14.7 $\pm$ 0.4 & 29.0 $\pm$ 0.6 & 44.8 $\pm$ 0.8 & 24.8 $\pm$ 0.5 \\
2614.5 & $^{208}$Tl       & 16.2 $\pm$ 0.2 & 17.4 $\pm$ 0.2 & 19.1 $\pm$ 0.3 & 28.5 $\pm$ 0.4 & 37.6 $\pm$ 0.5 & 48.6 $\pm$ 0.7 & 35.0 $\pm$ 0.4 \\
\hline
\multicolumn{2}{c}{\textbf{Cut value}} 
& $\mathbf{0.818}$ & $\mathbf{0.797}$ & $\mathbf{0.859}$ & $\mathbf{0.522}$ & $\mathbf{0.459}$ & $\mathbf{0.674}$ & $\mathbf{6.31}$ \\
\multicolumn{2}{c}{\textbf{$\alpha$ rejection factor}} 
& $\mathbf{> 2.62 \times10^4}$ & $\mathbf{> 2.71\times10^4}$ & $\mathbf{> 2.74\times10^4}$ & $\mathbf{84}$ & $\mathbf{96}$ & $\mathbf{2.89\times10^2}$ & $\mathbf{54}$ \\
\hline
\end{tabular}
\end{table*}

\begin{figure}
\centering
\includegraphics[width=0.45\textwidth]{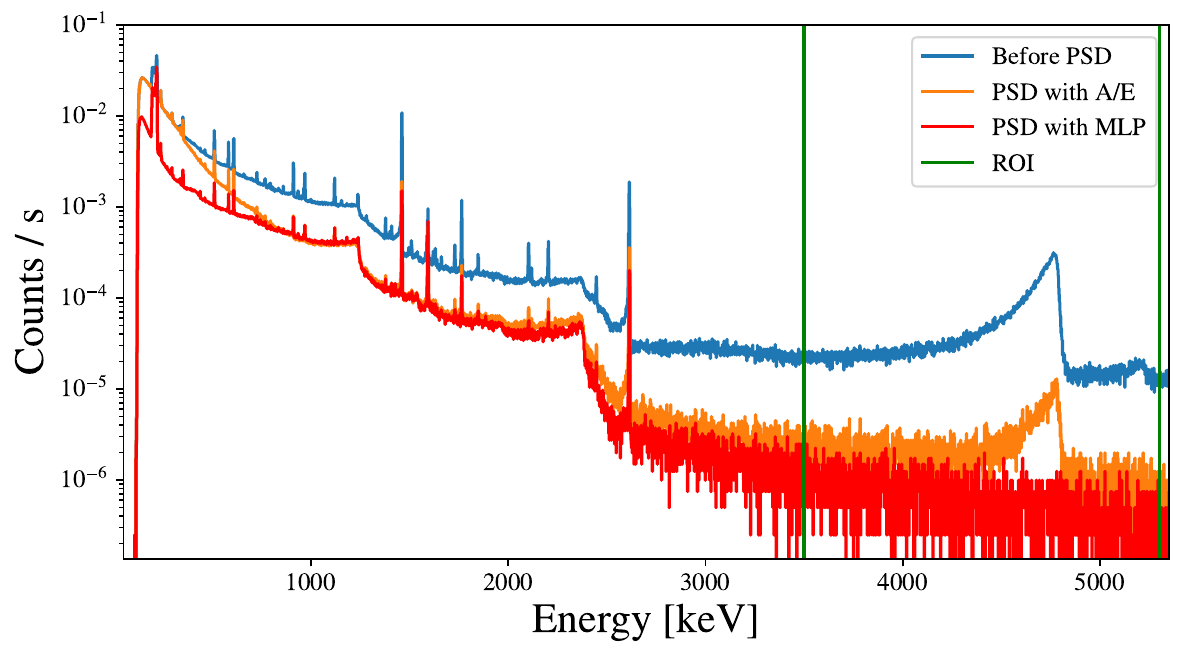}
\includegraphics[width=0.45\textwidth]{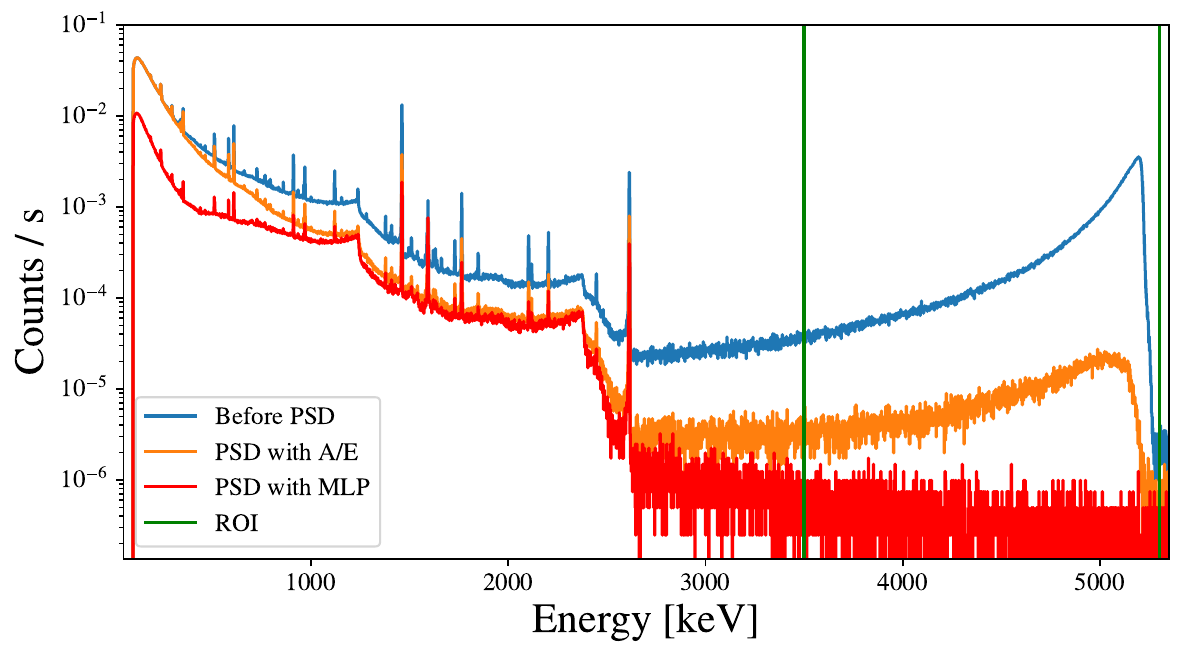}
\caption{Spectra of the low statistics (top) and high statistics (bottom) physics runs before and after PSD. A comparison is present between the $A/E$ (orange) and MLP (red) classifiers. In both cases, after PSD with MLP the peaks and tails from both polonium isotopes are strongly reduced, while after PSD with $A/E$ case the residual $\alpha$ events remain visible.}
\label{fig:spectra_aft} 
\end{figure}

The results tables (Tab. \ref{tab:results_HS_MLP} and \ref{sec:results_tables}) show that the MLP configurations TC2 and TC3 provide the best overall performance for both $\gamma$ and $\alpha$ rays in the HS and LS runs, with very similar results. Notably, the configurations that perform best in rejecting MSE gamma events also tend to perform best in rejecting alpha events.
Among the configurations with the best overall performance, TC2 was selected for the HS run and TC3 for the LS run, based on their slightly better general results in their respective datasets. For the HS run, TC2 gives DEP survival fractions above 80\%, MSE survival fractions below 20\%, and a 90\% C.L. limit on the $\alpha$-rejection factor of $2.71 \times 10^{4}$. For the LS run, TC3 provides a 90\% C.L. limit on the $\alpha$-rejection factor of $1.1 \times 10^{3}$. The final spectra of the physics runs for both the LS and HS campaigns, obtained using the TC3 and TC2 configurations of the MLP, respectively, are shown in Fig.~\ref{fig:spectra_aft}.
The peaks and tails originating from $^{209}\mathrm{Po}$ and $^{210}\mathrm{Po}$ are strongly suppressed. The remaining counts are consistent with residual muon background, as confirmed by the statistical analysis. The spectrum of the HS physics run before and after PSD with the $A/E$ classifier is also shown in Fig.~\ref{fig:spectra_aft} for comparison. Final results for the HS campaign using the MLP and $A/E$ methods are reported in Tab.~\ref{tab:results_HS_MLP}; additional results are provided in ~\ref{sec:results_tables}.
Configurations trained on the DEP (TC1, TC2, TC3) clearly outperform those trained on the CE (TC4, TC5, TC6) in rejecting both $\gamma$ and $\alpha$ backgrounds. This is due to the lower purity of the CE-based training samples. In general, the MLP performs better than both PLK and $A/E$ in rejecting the undesired $\gamma$ events. Comparing HS and LS runs, the PSD performance on the $\gamma$ background appears to have slightly worsened after the replacement of the JFET. However, the survival fractions for both runs are consistent with those reported in \cite{Jany2021} for the same detector.
Regarding the $\alpha$ background, both MLP and PLK classifiers perform very well, with the MLP achieving the best limits, because it yields fewer remaining events. Although the upper $A/E$ cut can reject a large fraction of $\alpha$ events, with a rejection factor of $\sim$50, this suppression is significantly weaker than that achieved by the competing methods, and the polonium peaks remain visible after the PSD cut in both the HS and LS cases, as it could be noticed from Fig. \ref{fig:spectra_aft}. While lowering the upper $A/E$ cut would further suppress $\alpha$ events, it would also lead to an unacceptably low survival fraction for the signal proxy.
These results demonstrate that both $\alpha$ and $\gamma$ backgrounds can be efficiently suppressed using a single pulse shape discrimination cut, while preserving the signal proxy with high efficiency.

\section{Conclusions and Outlook}
\label{sec:conclusions}
The research presented in this paper investigated the feasibility of rejecting both MSE $\gamma$ and $\alpha$ events in HPGe detectors using a classifier trained solely on $\gamma$ events. Two machine learning methods, MLP and PLK, were tested, with $A/E$ used for comparison. Both methods preserved a large fraction of the signal proxy while strongly reducing MSE $\gamma$ events and the $\alpha$ rays, showing better performance than the benchmark $A/E$ method. The MLP provided the best overall performance, achieving a stronger suppression of MSE events and providing stronger lower limits on the $\alpha$-rejection factor. For these reasons, the MLP approach is preferred. The MLP architecture used in this study represents one possible configuration chosen empirically. Although the hyperparameters were optimized via grid search, a systematic exploration of alternative architectures was not performed and may yield different results.
The sensitivity of the present study is limited by statistics and by the residual muon background. To extend the study, we propose the use of a stronger $\alpha$ source with several distinct alpha peaks, such as $^{226}\mathrm{Ra}$, and the repetition of the measurement campaign in an underground laboratory to reduce the muon component. Using the same detector described in this work, we measured the muon flux in the Książ Underground Laboratory ($\sim$ 270~m.w.e.) and found it to be a factor of 35 lower than at the surface. Repeating the same experiment underground would therefore likely lead to a final number of surviving events of $O(10)$, corresponding to a rejection factor in the order of $10^{5}$. Although this study focused on a BEGe-like detector, similar PSD performance is expected for other p-type point contact geometries, including the inverted coaxial detectors used in LEGEND. This expectation is motivated by previous experience with SSE and MSE discrimination in GERDA, MAJORANA DEMONSTRATOR and LEGEND, which suggests comparable PSD characteristics across these detector types. Nevertheless, the dependence of the present results on detector geometry should be verified experimentally, and a dedicated measurement campaign with a LEGEND-like inverted coaxial detector \cite{legend2025first} is therefore planned.

\begin{acknowledgements}
This work was supported by the Polish National Science Center (Grant Nos. UMO-2020/37/B/ST2/03905 and UMO-2025/57/N/ST2/04620). 
 
The research for this publication has been supported by the budget of Anthropocene Priority Research Area
(Earth System Science Core Facility Flagship Project) under the Strategic Programme Excellence Initiative at the Jagiellonian University.

We gratefully acknowledge Poland's high-performance Infrastructure PLGrid ACK Cyfronet AGH for providing computer facilities and support within computational grant No. PLG/2024/017397. This research used resources provided by National Energy Research Scientific Computing Center (NERSC), a U.S. Department of Energy Office of Science User Facility at Lawrence Berkeley National Laboratory, and the Oak Ridge Leadership Computing Facility at Oak Ridge National Laboratory.
\end{acknowledgements}

\bibliographystyle{spphys}       

\bibliography{biblio}   

\newpage
\appendix
\section{\label{sec:source_po209} Preparation of Po sources}
\begin{figure}[ht]
\centering
\includegraphics[width=0.45\textwidth]{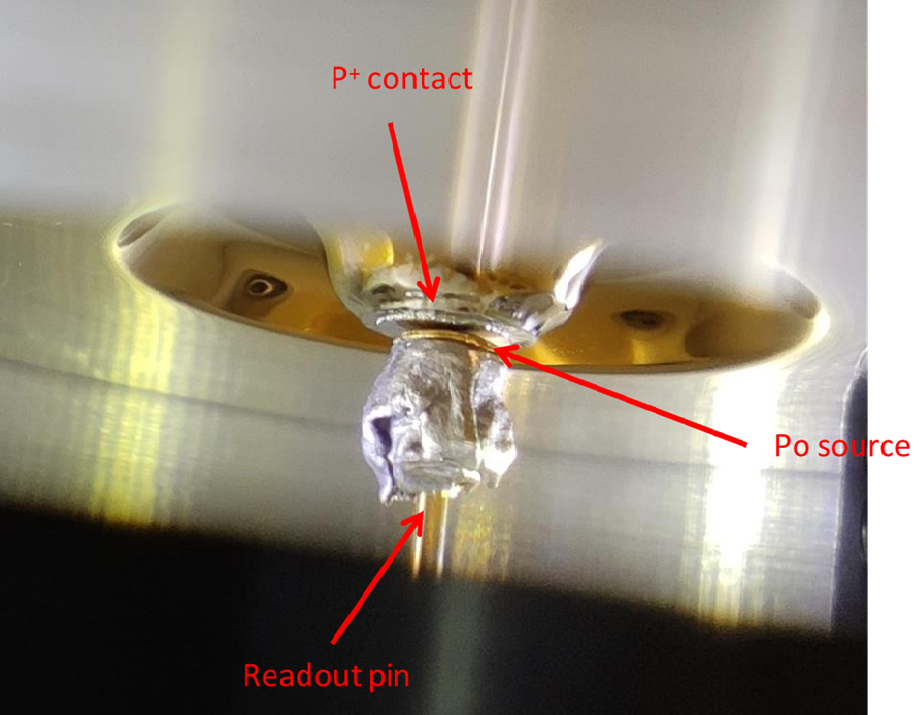}
\caption{Gold foil with deposited Po placed between the p+ contact of the detector and the readout pin.}
\label{fig:source}
\end{figure}

The most widely used approach for the preparation of Po sources is spontaneous deposition onto a silver disc, as described by Flynn \cite{flynn1968determination}. Polonium can also be relatively easily deposited on other noble metals, such as Pt or Pd, from diluted hydrochloric acid solutions in the presence of reducing agents; however, plating on gold requires a more complex procedure. Gold, being a good electrical conductor and a soft metal, was chosen as a plating material. Taking into account that, after attaching it to an indium foil, it will be placed between the p+ contact of the HPGe detector and the readout pin (see Fig. \ref{fig:source}). Po sources were prepared using a modified version of the procedure described by Erbacher \cite{erbacher1932neuartige}. In this method, polonium can be deposited on gold foil from a 0.9 M hydrochloric acid solution in the presence of thiourea, forming a stable gold-thiourea complex with a potential of +0.20 V. This enables the deposition of ${}^{209}\textrm{Po}$, for which the electrode potentials in HCl solution are +0.6 V for ${}\textrm{Po}^{0}/\textrm{Po}^{+2}$ and +0.8 V for ${}\textrm{Po}^{0}/\textrm{Po}^{+4}$, respectively. To increase the deposition efficiency, the method was modified by applying a potential of 5 V to the gold foil. Polonium was deposited on a gold foil disc with a diameter of 4 mm. Before deposition, the disc was sliced and immersed in the solution to maximize the polonium activity on a limited surface area.

The first source (LS run) was prepared by applying ${}^{209}\textrm{Po}$ taken from a standard NIST solution of Polonium Chloride (2 M HCl). 

The source for the HS run was based on ${}^{210}\textrm{Po}$ separated from normal lead (containing about 500 $\mathrm{Bq}$/kg of ${}^{210}\textrm{Pb}$). A few grams of lead were mechanically cleaned to remove potential surface oxidation or other contamination. The lead sample was then dissolved in 20 mL of concentrated nitric acid (14 - 15 M) at high temperature (150 - 220 $^\circ$C, solution in a beaker placed on a hot plate and constantly stirred). In the next steps,
nitric acid was gradually evaporated and replaced by hydrochloric acid: lead chloride precipitated, while polonium remained in solution. After the solution was cooled, it was separated from the precipitate by centrifugation. The supernatant was transferred to a 250 mL beaker and the solution was evaporated on a hot plate until reduced to a small volume ($\sim$5 mL). Next, 6 mL of concentrated hydrochloric acid was added and evaporated almost to dryness. This step was repeated three times. Finally, 6~mL of concentrated hydrochloric acid, 144~mL of water, and 0.5~g of hydrazine dihydrochloride were added to the residue. The ${}^{210}\textrm{Po}$ deposition from this solution was carried out as described above.

\section{Results Tables}
\label{sec:results_tables}
\begin{table*}[htbp]
\centering
\caption{Survival fractions for low statistics dataset (MLP configurations TC1--TC6 and A/E benchmark).}
\label{tab:results_LS_MLP}
\begin{tabular}{c c *{6}{c} c}
\hline
\multirow{2}{*}{Energy [keV]} &
\multirow{2}{*}{Isotope} &
\multicolumn{7}{c}{Survival Fraction [\%]} \\
\cline{3-9}
 & & TC1 & TC2 & TC3 & TC4 & TC5 & TC6 & A/E \\
\hline
727.3  & $^{212}$Bi       & 21.6 $\pm$ 0.4 & 21.7 $\pm$ 0.4 & 21.6 $\pm$ 0.4 & 53.6 $\pm$ 0.9 & 40.8 $\pm$ 0.7 & 26.7 $\pm$ 0.4 & 33.2 $\pm$ 0.6 \\
763.0  & $^{208}$Tl       & 20.4 $\pm$ 1.6 & 21.3 $\pm$ 1.4 & 21.4 $\pm$ 1.4 & 49.7 $\pm$ 2.4 & 39.6 $\pm$ 1.9 & 28.2 $\pm$ 1.5 & 31.6 $\pm$ 1.7 \\
785.0  & $^{212}$Bi       & 20.1 $\pm$ 1.1 & 21.0 $\pm$ 1.0 & 20.6 $\pm$ 1.0 & 49.6 $\pm$ 1.5 & 41.2 $\pm$ 1.4 & 28.2 $\pm$ 1.1 & 28.8 $\pm$ 1.1 \\
860.6  & $^{208}$Tl       & 19.9 $\pm$ 0.4 & 20.3 $\pm$ 0.4 & 19.9 $\pm$ 0.4 & 50.0 $\pm$ 1.0 & 41.3 $\pm$ 0.8 & 27.1 $\pm$ 0.5 & 26.0 $\pm$ 0.5 \\
1078   & $^{212}$Bi       & 18.4 $\pm$ 2.7 & 18.5 $\pm$ 2.5 & 17.6 $\pm$ 2.4 & 43.4 $\pm$ 3.6 & 39.4 $\pm$ 3.1 & 24.1 $\pm$ 2.5 & 21.6 $\pm$ 2.5 \\
1093   & $^{208}$Tl       & < 5            & 10.8 $\pm$ 6.6 & 8.8 $\pm$ 5.3  & 55.9 $\pm$ 7.3 & 33.2 $\pm$ 5.9 & 14.0 $\pm$ 5.0 & 7.0 $\pm$ 5.5 \\
1512   & $^{212}$Bi       & 16.0 $\pm$ 4.5 & 17.1 $\pm$ 4.5 & 16.7 $\pm$ 4.5 & 39.3 $\pm$ 6.4 & 42.4 $\pm$ 7.0 & 26.0 $\pm$ 5.5 & 25.0 $\pm$ 4.6 \\
\textbf{1592} & \textbf{$^{208}$Tl (DEP)} 
        & $\mathbf{92.3 \pm 1.3}$ & $\mathbf{92.8 \pm 1.3}$ & $\mathbf{92.1 \pm 1.3}$ & $\mathbf{94.0 \pm 1.3}$ & $\mathbf{86.2 \pm 1.3}$ & $\mathbf{94.2 \pm 1.3}$ & $\mathbf{85.2 \pm 1.2}$ \\
1620.5 & $^{212}$Bi       & 15.7 $\pm$ 1.2 & 15.4 $\pm$ 1.2 & 15.2 $\pm$ 1.2 & 33.4 $\pm$ 1.9 & 40.0 $\pm$ 2.0 & 22.8 $\pm$ 1.4 & 22.0 $\pm$ 1.5 \\
2103   & $^{208}$Tl (SEP) & 9.1 $\pm$ 0.7  & 9.9 $\pm$ 0.7  & 9.3 $\pm$ 0.7  & 17.4 $\pm$ 1.0 & 36.7 $\pm$ 1.3 & 16.3 $\pm$ 0.8 & 15.5 $\pm$ 0.7 \\
2614.5 & $^{208}$Tl       & 13.6 $\pm$ 0.3 & 14.0 $\pm$ 0.3 & 13.1 $\pm$ 0.3 & 34.3 $\pm$ 0.8 & 58.3 $\pm$ 1.4 & 23.2 $\pm$ 0.5 & 24.3 $\pm$ 0.5 \\
\hline
\multicolumn{2}{c}{\textbf{Cut value}} 
& $\mathbf{0.664}$ & $\mathbf{0.572}$ & $\mathbf{0.582}$ & $\mathbf{0.602}$ & $\mathbf{0.346}$ & $\mathbf{0.541}$ & $\mathbf{2.45}$ \\
\multicolumn{2}{c}{\textbf{$\alpha$ rejection factor}} 
& $\mathbf{> 7.5 \times 10^2}$ & $\mathbf{> 1.0 \times 10^3}$ & $\mathbf{> 1.1 \times 10^3}$ & $\mathbf{11}$ & $\mathbf{3}$ & $\mathbf{> 7.5 \times 10^2}$ & $\mathbf{30}$ \\
\hline
\end{tabular}
\end{table*}

\begin{table*}[htbp]
\centering
\caption{Survival fractions for low statistics dataset (PLK configurations TC1--TC6 and A/E benchmark).}
\label{tab:results_LS_PLK}
\begin{tabular}{c c *{6}{c} c}
\hline
\multirow{2}{*}{Energy [keV]} &
\multirow{2}{*}{Isotope} &
\multicolumn{7}{c}{Survival Fraction [\%]} \\
\cline{3-9}
 & & TC1 & TC2 & TC3 & TC4 & TC5 & TC6 & A/E \\
\hline
727.3  & $^{212}$Bi       & 18.7 $\pm$ 0.3 & 10.4 $\pm$ 0.2 & 10.6 $\pm$ 0.2 & 15.4 $\pm$ 0.3 & 9.1 $\pm$ 0.2  & 18.2 $\pm$ 0.3 & 33.2 $\pm$ 0.6 \\
763.0  & $^{208}$Tl       & 17 $\pm$ 2     & 12 $\pm$ 1     & 12 $\pm$ 1     & 16 $\pm$ 1     & 10.2 $\pm$ 0.9 & 20 $\pm$ 1     & 31.6 $\pm$ 1.7 \\
785.0  & $^{212}$Bi       & 17 $\pm$ 1     & 11.3 $\pm$ 0.6 & 11.7 $\pm$ 0.7 & 15.1 $\pm$ 0.9 & 11.2 $\pm$ 0.6 & 20.1 $\pm$ 0.9 & 28.8 $\pm$ 1.1 \\
860.6  & $^{208}$Tl       & 17.1 $\pm$ 0.3 & 14.1 $\pm$ 0.3 & 13.0 $\pm$ 0.3 & 17.1 $\pm$ 0.4 & 14.1 $\pm$ 0.3 & 21.1 $\pm$ 0.4 & 26.0 $\pm$ 0.5 \\
1078   & $^{212}$Bi       & 20 $\pm$ 3     & 20 $\pm$ 2     & 19 $\pm$ 2     & 22 $\pm$ 3     & 22 $\pm$ 2     & 25 $\pm$ 3     & 21.6 $\pm$ 2.5 \\
1093   & $^{208}$Tl       & 5 $\pm$ 5      & 4 $\pm$ 4      & < 5            & 8 $\pm$ 5      & 10 $\pm$ 3     & 14 $\pm$ 5     & 7.0 $\pm$ 5.5 \\
1512   & $^{212}$Bi       & 23 $\pm$ 5     & 28 $\pm$ 5     & 26 $\pm$ 5     & 30 $\pm$ 5     & 33 $\pm$ 5     & 33 $\pm$ 6     & 25.0 $\pm$ 4.6 \\
\textbf{1592} & \textbf{$^{208}$Tl (DEP)} 
        & $\mathbf{90.9 \pm 1.3}$ & $\mathbf{91.9 \pm 1.3}$ & $\mathbf{90.5 \pm 1.3}$ & $\mathbf{86.3 \pm 1.2}$ & $\mathbf{86.7 \pm 1.2}$ & $\mathbf{91.9 \pm 1.3}$ & $\mathbf{85.2 \pm 1.2}$ \\
1620.5 & $^{212}$Bi       & 23.3 $\pm$ 1.4 & 27.5 $\pm$ 1.5 & 23.3 $\pm$ 1.4 & 28.7 $\pm$ 1.7 & 35 $\pm$ 2     & 33 $\pm$ 2     & 22.0 $\pm$ 1.5 \\
2103   & $^{208}$Tl (SEP) & 15.9 $\pm$ 0.7 & 20.3 $\pm$ 0.9 & 16.5 $\pm$ 0.8 & 21.2 $\pm$ 0.9 & 29 $\pm$ 1     & 25.7 $\pm$ 1.0 & 15.5 $\pm$ 0.7 \\
2614.5 & $^{208}$Tl       & 24.4 $\pm$ 0.5 & 29.9 $\pm$ 0.6 & 24.9 $\pm$ 0.5 & 32.6 $\pm$ 0.7 & 41.8 $\pm$ 0.9 & 37.0 $\pm$ 0.8 & 24.3 $\pm$ 0.5 \\
\hline
\multicolumn{2}{c}{\textbf{Cut value}} 
& $\mathbf{0.853}$ & $\mathbf{0.638}$ & $\mathbf{0.885}$ & $\mathbf{0.681}$ & $\mathbf{0.595}$ & $\mathbf{0.735}$ & $\mathbf{2.45}$ \\
\multicolumn{2}{c}{\textbf{$\alpha$ rejection factor}} 
& $\mathbf{>5.5\times10^{2}}$ 
& $\mathbf{>7.1\times10^{2}}$ 
& $\mathbf{>6.4\times10^{2}}$ 
& $\mathbf{8.1}$ 
& $\mathbf{>2.5\times10^{2}}$ 
& $\mathbf{>6.4\times10^{2}}$ 
& $\mathbf{30}$ \\
\hline
\end{tabular}
\end{table*}

\begin{table*}[htbp]
\centering
\caption{Survival fractions for high statistics dataset (PLK configurations TC1--TC6 and A/E benchmark).}
\label{tab:results_HS_PLK}
\begin{tabular}{c c *{6}{c} c}
\hline
\multirow{2}{*}{Energy [keV]} &
\multirow{2}{*}{Isotope} &
\multicolumn{7}{c}{Survival Fraction [\%]} \\
\cline{3-9}
 & & TC1 & TC2 & TC3 & TC4 & TC5 & TC6 & A/E \\
\hline
727.3  & $^{212}$Bi       & 22.7 $\pm$ 0.2 & 36.7 $\pm$ 0.4 & 40.6 $\pm$ 0.4 & 25.5 $\pm$ 0.3 & 47.0 $\pm$ 0.5 & 60.5 $\pm$ 0.7 & 40.5 $\pm$ 0.4 \\
763.0  & $^{208}$Tl       & 24.7 $\pm$ 0.6 & 38.1 $\pm$ 0.8 & 41.1 $\pm$ 0.8 & 27.0 $\pm$ 0.7 & 48.3 $\pm$ 0.8 & 60.9 $\pm$ 1.0 & 40.2 $\pm$ 0.8 \\
785.0  & $^{212}$Bi       & 24.9 $\pm$ 0.4 & 37.7 $\pm$ 0.6 & 40.1 $\pm$ 0.6 & 26.2 $\pm$ 0.4 & 48.7 $\pm$ 0.6 & 60.7 $\pm$ 0.8 & 38.8 $\pm$ 0.5 \\
860.6  & $^{208}$Tl       & 24.2 $\pm$ 0.2 & 39.1 $\pm$ 0.3 & 40.0 $\pm$ 0.4 & 27.3 $\pm$ 0.2 & 51.4 $\pm$ 0.4 & 60.6 $\pm$ 0.5 & 37.1 $\pm$ 0.3 \\
1078   & $^{212}$Bi       & 25.7 $\pm$ 1.1 & 39.9 $\pm$ 1.4 & 39.7 $\pm$ 1.3 & 30.2 $\pm$ 1.3 & 53.9 $\pm$ 1.6 & 59.7 $\pm$ 1.6 & 33.6 $\pm$ 1.4 \\
1093   & $^{208}$Tl       & 24.6 $\pm$ 3.1 & 35.0 $\pm$ 3.5 & 34.2 $\pm$ 3.6 & 29.4 $\pm$ 3.4 & 46.7 $\pm$ 4.1 & 54.0 $\pm$ 4.4 & 27.3 $\pm$ 3.3 \\
1512   & $^{212}$Bi       & 26.9 $\pm$ 2.7 & 40.4 $\pm$ 2.8 & 39.5 $\pm$ 2.7 & 33.4 $\pm$ 3.1 & 54.7 $\pm$ 2.9 & 57.3 $\pm$ 3.2 & 31.9 $\pm$ 2.4 \\
\textbf{1592} & \textbf{$^{208}$Tl (DEP)} 
        & $\mathbf{81.6 \pm 0.8}$ & $\mathbf{80.1 \pm 0.8}$ & $\mathbf{78.2 \pm 0.8}$ & $\mathbf{79.5 \pm 0.8}$ & $\mathbf{88.5 \pm 0.9}$ & $\mathbf{88.3 \pm 0.9}$ & $\mathbf{73.6 \pm 0.7}$ \\
1620.5 & $^{212}$Bi       & 26.9 $\pm$ 0.8 & 41.3 $\pm$ 0.9 & 40.0 $\pm$ 1.0 & 34.6 $\pm$ 0.8 & 56.6 $\pm$ 1.1 & 59.2 $\pm$ 1.1 & 33.0 $\pm$ 0.8 \\
2103   & $^{208}$Tl (SEP) & 18.2 $\pm$ 0.4 & 37.3 $\pm$ 0.6 & 34.7 $\pm$ 0.6 & 30.2 $\pm$ 0.5 & 54.7 $\pm$ 0.9 & 55.8 $\pm$ 0.9 & 24.8 $\pm$ 0.5 \\
2614.5 & $^{208}$Tl       & 25.4 $\pm$ 0.3 & 43.2 $\pm$ 0.6 & 41.2 $\pm$ 0.5 & 40.6 $\pm$ 0.5 & 59.6 $\pm$ 0.7 & 61.4 $\pm$ 0.8 & 35.0 $\pm$ 0.4 \\
\hline
\multicolumn{2}{c}{\textbf{Cut value}} 
& $\mathbf{0.821}$ & $\mathbf{0.931}$ & $\mathbf{0.875}$ & $\mathbf{0.778}$ & $\mathbf{0.370}$ & $\mathbf{0.690}$ & $\mathbf{6.31}$ \\
\multicolumn{2}{c}{\textbf{$\alpha$ rejection factor}} 
& $\mathbf{>1.88\times10^{4}}$ 
& $\mathbf{>1.51\times10^{4}}$ 
& $\mathbf{>1.65\times10^{4}}$ 
& $\mathbf{2.4\times10^{2}}$ 
& $\mathbf{67}$ 
& $\mathbf{>1.32\times10^{4}}$ 
& $\mathbf{54}$ \\
\hline
\end{tabular}
\end{table*}

\end{document}